# ON THE FLOW OF OLDROYD-B FLUIDS WITH FRACTIONAL DERIVATIVES OVER A PLATE THAT APPLIES SHEAR STRESS TO THE FLUID

**Azhar Ali Zafar[1,*], Constantin Fetecau[2,3], Itrat Abbas Mirza[1]**

1. Abdus Salam School of Mathematical Sciences, GC University Lahore, Pakistan
2. Technical University of Iasi, Romania
3. Academy of Romanian Scientists, 050094 Bucuresti, Romania.

**Abstract**

The motion of incompressible fractional Oldroyd-B fluids between two parallel walls perpendicular to a plate that applies time-dependent shear stresses to the fluid is studied by means of integral transforms. In the special cases of Newtonian and second grade fluids, these shear stresses reduces to $fH(t)\sin(\omega t)$ or $fH(t)\cos(\omega t)$. General solutions for velocity are presented as a sum of Newtonian solutions and the corresponding non-Newtonian contributions. They reduce to the similar solutions corresponding to the motion over an infinite plate if the distance between walls tend to infinity and can be easy particularized to give the similar solutions for ordinary and fractional Maxwell or second grade fluids performing the same motions. As a check of general results some known solutions from the literature are recovered as limiting cases. Finally, the influence of fractional parameters on the fluid motion and the distance between walls for which the measured value of the velocity in the middle of the channel is unaffected by their presence (more exactly, it is equal to the velocity corresponding to the motion over an infinite plate) are graphically determined.

## 1. Introduction

Historically, the Oldroyd-B model is regarded as an extension of the upper convected Maxwell model [1] that has gained a special status amongst the rate type fluids. It represents one of the simplest constitutive models, capable of describing the visco-elastic behavior of dilute polymeric solutions under general flow conditions. Oldroyd-B fluids store energy like the linearized elastic solids, their dissipation being due to two dissipative mechanisms that implies that they arise from a mixture of two viscous fluids. Although the constitutive relations of Oldroyd-B fluids are relatively simple, the dynamics of several flow models is complicated enough to offer challenging numerical simulations to the mathematicians and numerical analysts. To illustrate the complexity involved in using such a model, we recommend the work of Rajagopal [2, Sec. 4]. Some existence results regarding flows of Oldroyd-B fluids have been extablished by Guillope and Sout [3].

The motion of Oldroyd-B fluids over an infinite plate as well as that between two parallel walls perpendicular to a plate was extensively studied in the literature (see for instance [4-12] and therein references). Such a motion can be induced by the plate that is moving in its plane or applies a shear stress to the fluid and some exact solutions have been also extended to Oldroyd-B fluids with fractional derivatives [13-16]. However, all these solutions correspond to motions due to the plate that is moving in its plane and velocity is given on the boundary. In some problems, what is specified is the force applied on the boundary [17-19]. Consequently, contrary to what is usually assumed, the force with which the plate is moved can be prescribed. To reiterate, in Newtonian mechanics force is the cause and kinematic is the effect (see Rajagopal [20] for a detailed discussion of the same). Furthermore the "no slip" boundary condition may not be necessarily applicable to flows of polymeric fluids that can slide on the boundary.

\* Corresponding author: e-mail: azharalizafar@gmail.com

On the other hand, the fractional calculus is an efficient tool and suitable frame-work within which useful generalizations of various classical physical concepts are already obtained. It was successfully used in describing the visco-elasicity [21] and the area of its applications is wide enough. Germant [22] seems to be the first who proposed the use of fractional derivatives in visco-elasticity and the interest for visco-elastic fluids with fractional derivatives came from practical problems. The first objective law which characterizes a fluid with fractional derivatives is that of Palade et al. [23, Eq. (16)]. Their constitutive relation under linearization is equivalent to the one-dimentional equality $\tau + \lambda D^\alpha \tau = \mu D^\beta \gamma$ proposed by Makris et al. [24] for the fractional Maxwell model. It is worth pointing out that mechanical properties predicted by means of this equation, where $\tau$ and $\gamma$ are the shear stress and strain and $D^\alpha$ is a fractional derivative operator with respect to time, were in excellent agreements with experimental results.

Based on the above remarks, our interest here is to extend some known results [25] to fluids with fractional derivatives. Consequently, we study the motion of an incompressible fractional Oldroyd-B fluid between two parallel walls perpendicular to a plate . The motion of the fluid is due to the plate that after time $t = 0^+$ applies a time-dependent shear stress to the fluid. In the special cases of second grade and Newtonian fluids, this shear stress reduces to an oscillating one and known results from the literature are recovered as limiting cases. Furthermore, the similar solutions corresponding to the motion over an infinite plate are also attained as special cases of general solution. They satisfy all imposed initial and boundary conditions and can easy be reduced to the similar solutions for fractional or ordinary Maxwell, second grade and Newtonian fluids. Finally, the influence of fractional parameters on the fluid motion as well as the distance between walls for which their presence can be neglected is graphically determined.

## 2. Governing equations

In the following we propose to study the unidirectional motion of fractional Oldroyd-B fluids whose velocity field $\mathbf{v}$ is of the form
$$\mathbf{v} = \mathbf{v}(y,z,t) = u(y,z,t)\mathbf{i}, \tag{1}$$
where $\mathbf{i}$ is the unit vector along the $x$-direction of a fixed Cartesian coordinate system $x$, $y$ and $z$. For such a motion the constraint of incompressibility is automatically satisfied. We also assume that the extra-stress tensor $\mathbf{S}$, as well as the velocity $\mathbf{v}$, depends of $y$, $z$ and $t$ only. By substituting Eq. (1) into the constitutive equations of an Oldroyd-B fluid and assuming that the fluid is at rest up to the moment $t = 0$, we obtain the relevant equations [9]
$$(1+\lambda\frac{\partial}{\partial t})\tau_1(y,z,t) = \mu(1+\lambda_r\frac{\partial}{\partial t})\frac{\partial u(y,z,t)}{\partial y}, \quad (1+\lambda\frac{\partial}{\partial t})\tau_2(y,z,t) = \mu(1+\lambda_r\frac{\partial}{\partial t})\frac{\partial u(y,z,t)}{\partial z}. \tag{2}$$
Here $\tau_1(y,z,t) = S_{xy}(y,z,t)$ and $\tau_2(y,z,t) = S_{xz}(y,z,t)$ are the non-trivial shear stresses, $\mu$ is the dynamic viscocity of the fluid while $\lambda$ and $\lambda_r$ are relaxation, respectively retardation times.

By neglecting the body forces and in the absence of a pressure gradient in the flow direction, the balance of linear momentum leads to the significant equation

$$\frac{\partial \tau_1(y,z,t)}{\partial y} + \frac{\partial \tau_2(y,z,t)}{\partial z} = \rho \frac{\partial u(y,z,t)}{\partial y}, \quad (3)$$

where $\rho$ is the constant density of the fluid. Eliminating $\tau_1(y,z,t)$ and $\tau_2(y,z,t)$ between Eqs. (2) and (3), the governing equation

$$(1+\lambda \frac{\partial}{\partial t})\frac{\partial u(y,z,t)}{\partial t} = \nu(1+\lambda_r \frac{\partial}{\partial t})\left[\frac{\partial^2}{\partial y^2} + \frac{\partial^2}{\partial z^2}\right]u(y,z,t), \quad (4)$$

for velocity is obtained. Here $\nu = \mu/\rho$ is the kinematic viscosity of the fluid.

In order to determine the velocity field corresponding to a motion problem with velocity on the boundary, the partial differential equation (4) is sufficient. However, for a shear stress boundary value problem both equations (2) and (4) are necessary. As usual, the governing equations corresponding to such a motion of fractional Oldroyd-B fluids, namely

$$(1+\lambda^\alpha D_t^\alpha)\tau_1(y,z,t) = \mu(1+\lambda_r^\beta D_t^\beta)\frac{\partial u(y,z,t)}{\partial y}, \quad (5)$$

$$(1+\lambda^\alpha D_t^\alpha)\tau_2(y,z,t) = \mu(1+\lambda_r^\beta D_t^\beta)\frac{\partial u(y,z,t)}{\partial z}, \quad (6)$$

$$(1+\lambda^\alpha D_t^\alpha)\frac{\partial u(y,z,t)}{\partial t} = \nu(1+\lambda_r^\beta D_t^\beta)\left(\frac{\partial^2}{\partial y^2} + \frac{\partial^2}{\partial z^2}\right)u(y,z,t), \quad (7)$$

are derived from Eqs. (2) and (4) by substituting the inner time derivatives by the Caputo derivative operator [26, 27]

$$D_t^p f(t) = \frac{1}{\Gamma(1-p)}\int_0^t \frac{f'(\tau)}{(t-\tau)^p}d\tau, \; 0 < p < 1, \; D_t^1 f(t) = f'(t). \quad (8)$$

Into above relations $\alpha \geq \beta$ [28] and $\Gamma(.)$ is the Gamma function. For $\alpha = \beta = 1$ the fractional Oldroyd-B model reduces to the usual Oldroyd-B model and for $\beta = 1$ and $\lambda_r = 0$ the fractional Maxwell model is obtained.

### 3. Wording and solution of the problem

Consider an incompressible fractional Oldroyd-B fluid between two parallel walls (in the planes $z = 0$ and $z = d$) perpendicular to a plate situated at the plane $y = 0$. After time $t = 0$, the plate is set in motion due to a time dependent shear stress according to the governing equation (5). It is chosen so that for second grade and Newtonian fluids to become [29, 30]

$$\tau_1(0,z,t) = f\,H(t)\sin(\omega t) \text{ and } \tau_1(0,z,t) = f\,H(t)\cos(\omega t), \quad (9)$$

where $f$ is a constant, $\omega$ is the frequency of the oscillations and $H(\cdot)$ is the Heaviside unit step function. Due to the shear the fluid is gradually moved and its velocity is of the form (1). The governing equation for velocity is given by the equality (7) and the appropriate initial and boundary conditions are

$$u(y,z,0) = 0, \; \frac{\partial u(y,z,t)}{\partial t}\Big|_{t=0} = 0; \quad y > 0, \quad (10)$$

$$(1+\lambda^\alpha D_t^\alpha)\tau_1(0,z,t) = \mu(1+\lambda_r^\beta D_t^\beta)\frac{\partial u(y,z,t)}{\partial y}\Big|_{y=0} \tag{11}$$

$$= f\,H(t)\sin(\omega t) \text{ or } f\,H(t)\cos(\omega t); \quad z\in(0,d),$$

$$u(y,0,t) = u(y,d,t) = 0;\quad y>0,\ t\geq 0, \tag{12}$$

$$u(y,z,t),\frac{\partial u(y,z,t)}{\partial y}\to 0 \text{ as } y\to\infty;\ z\in[0,d],\ t\geq 0. \tag{13}$$

Of course, in view of Eq. (11), after the moment $t=0^+$ the plate applies to the fluid shear stresses of the form

$$\tau_1(0,z,t) = \frac{f}{\lambda^\alpha}H(t)\int_0^t \sin(\omega(t-s))G_{\alpha,0,1}(-\frac{1}{\lambda^\alpha},s)ds, \tag{14}$$

or

$$\tau_1(0,z,t) = \frac{f}{\lambda^\alpha}H(t)\int_0^t \cos(\omega(t-s))G_{\alpha,0,1}(-\frac{1}{\lambda^\alpha},s)ds, \tag{15}$$

where the generalized function $G_{a,b,c}(d,t) = \sum_{j=0}^\infty \frac{d^j\Gamma(j+c)t^{a(j+c)-b-1}}{\Gamma(j+1)\Gamma(c)\Gamma(a(j+c)-b)}$ [31, pp. 14-15] is the inverse Laplace transform of $\frac{q^b}{(q^a-d)^c}$ if $\text{Re}(ac-b)>0$, $\text{Re}(q)>0$ and $\left|\frac{d}{q^a}\right|<1$. For $\alpha=\beta=1$, corresponding to ordinary Oldroyd-B fluids, Eqs. (14) and (15) take the simple forms [25, Eqs. (6) and (7)]

$$\tau_1(0,z,t) = f\,H(t)\frac{\lambda\omega}{1+\lambda^2\omega^2}\left\{\frac{1}{\lambda\omega}\sin(\omega t)-\cos(\omega t)+e^{\frac{-t}{\lambda}}\right\}, \tag{16}$$

or

$$\tau_1(0,z,t) = f\,H(t)\frac{\lambda\omega}{1+\lambda^2\omega^2}\left\{\sin(\omega t)+\frac{1}{\lambda\omega}\cos(\omega t)-\frac{1}{\lambda\omega}e^{\frac{-t}{\lambda}}\right\}. \tag{17}$$

For $\lambda\to 0$, Eqs. (16) and (17) tend to the equations (9) corresponding to sine and cosine shear stresses on the boundary. In the following the linear partial differential equation (7), together with the initial and boundary conditions (10) - (13), will be solved using Fourier and Laplace transforms. For simplicity, the obtained solutions will be refered as to sine and cosine solutions.

In order to solve both problems in the same time, let us denote by $u_c(y,z,t)$ and $u_s(y,z,t)$ the solutions corresponding to the cosine and sine oscillations and by

$$v(y,z,t) = u_c(y,z,t)+iu_s(y,z,t), \tag{18}$$

the complex velocity where $i$ is the imaginary unit. With a view to the previous relations, we attain to the next initial and boundary-value problem

$$(1+\lambda^\alpha D_t^\alpha)\frac{\partial v(y,z,t)}{\partial t} = \nu(1+\lambda_r^\beta D_t^\beta)\left(\frac{\partial^2}{\partial y^2}+\frac{\partial^2}{\partial z^2}\right)v(y,z,t);\ y,t>0, \tag{19}$$

$$v(y,z,0) = \frac{\partial v(y,z,t)}{\partial t}\Big|_{t=0} = 0;\ y>0,\ z\in[0,d], \tag{20}$$

$$\mu(1+\lambda_r^\beta D_t^\beta)\frac{\partial v(y,z,t)}{\partial y}\Big|_{y=0} = fH(t)e^{i\omega t}; \ z \in (0,d), \tag{21}$$

$$v(y,0,t) = v(y,d,t) = 0; \ y > 0, \ t \geq 0, \tag{22}$$

$$v(y,z,t), \ \frac{\partial v(y,z,t)}{\partial y} \to 0 \text{ as } y \to \infty; \ z \in [0,d], \ t \geq 0. \tag{23}$$

By multiplying Eq. (19) with $\sqrt{2/\pi}\cos(y\xi)\sin(\alpha_k z)$, where $\alpha_k = \frac{k\pi}{d}$, integrating the result with respect to $y$ and $z$ from 0 to $\infty$, respectively 0 to $d$ and using the boundary conditions (21) - (23), we find that

$$(1+\lambda^\alpha D_t^\alpha)\frac{\partial v_k(\xi,t)}{\partial t}+\nu(\xi^2+\alpha_k^2)(1+\lambda_r^\beta D_t^\beta)v_k(\xi,t) = \frac{f}{\rho}\sqrt{\frac{2}{\pi}}\left[\frac{(-1)^k-1}{\alpha_k}\right]e^{i\omega t}; t>0. \tag{24}$$

Of course in view of the initial conditions (20), the double Fourier cosine and sine transforms

$$v_k(\xi,t) = \sqrt{\frac{2}{\pi}}\int_0^\infty\int_0^d v(y,z,t)\cos(y\xi)\sin(\alpha_k z)\,dz\,dy, \ k = 1,2,3,....,$$

of $v(y,z,t)$ have to satisfy the initial conditions

$$v_k(\xi,0) = \frac{\partial v_k(\xi,t)}{\partial t}\Big|_{t=0} = 0 \text{ for } \xi > 0 \text{ and } k = 1,2,3,.... \tag{25}$$

Applying the Laplace transform to Eq. (24) and bearing in mind the initial conditions (25), we can obtain the image function $\overline{v}_k(\xi,q)$ of $v_k(\xi,t)$ in the form

$$\overline{v}_k(\xi,q) = \frac{f}{\rho}\sqrt{\frac{2}{\pi}}\left[\frac{(-1)^k-1}{\alpha_k}\right]\frac{1}{(q-i\omega)(\lambda^\alpha q^{\alpha+1}+q+\nu(\xi^2+\alpha_k^2)(1+\lambda_r^\beta q^\beta))}, \tag{26}$$

where $q$ is the Laplace transform parameter. In order to provide a suitable form of the velocity field, we rewrite Eq. (26) in the equivalent form

$$\overline{v}_k(\xi,q) = \frac{f}{\rho}\sqrt{\frac{2}{\pi}}\left[\frac{(-1)^k-1}{\alpha_k}\right](F_{1k}(\xi,q) - F_{2k}(\xi,q)F_{3k}(\xi,q)), \tag{27}$$

where $F_{1k}(\xi,q) = \dfrac{1}{(q-i\omega)(q+\nu(\xi^2+\alpha_k^2))}$, $F_{2k}(\xi,q) = q^\beta F_{1k}(\xi,q)$

and (28)

$$F_{3k}(\xi,q) = \frac{q^{\alpha-\beta}+\nu(\xi^2+\alpha_k^2)aq^{-1}}{(q^\alpha+\lambda^{-\alpha})+\lambda^{-\alpha}q^{-1}\nu(\xi^2+\alpha_k^2)+\nu(\xi^2+\alpha_k^2)aq^{\beta-1}} \text{ with } a = \frac{\lambda_r^\beta}{\lambda^\alpha}.$$

Finally applying the inverse Laplace transform to Eq. (27), inverting the result by means of inverse Fourier sine and cosine formulae [32] and using the convolution theorem, we attain for the complex velocity $v(y,z,t)$ the simple form

$$v(y,z,t) = -\frac{8f}{\rho\pi d}\sum_{k=1}^\infty\frac{\sin(\alpha_m z)}{\alpha_m}\int_0^\infty f_{1m}(\xi,t)\cos(y\xi)d\xi + $$
$$+\frac{8f}{\rho\pi d}\sum_{k=1}^\infty\frac{\sin(\alpha_m z)}{\alpha_m}\int_0^\infty\int_0^t f_{2m}(\xi,t-\tau)f_{3m}(\xi,\tau)d\tau d\xi, \tag{29}$$

where $m = 2k-1$ and $f_{1m}(\xi,t), f_{2m}(\xi,t)$ and $f_{3m}(\xi,t)$ are the inverse Laplace transforms of the functions $F_{1m}(\xi,q), F_{2m}(\xi,q)$ and $F_{3m}(\xi,q)$, respectively. By setting $d = 2h$ and changing the origin of the coordinate system at the middle of the channel (taking $z = z'+h$ and dropping out the prime notation), the complex velocity can be written in the equivalent form

$$v(y,z,t) = \frac{4f}{\rho\pi h}\sum_{k=1}^{\infty}\frac{(-1)^k \cos(\gamma_m z)}{\gamma_m}\int_0^{\infty} f_{1m}(\xi,t)\cos(y\xi)d\xi + \\ + \frac{4f}{\rho\pi h}\sum_{k=1}^{\infty}\frac{(-1)^{k+1}\cos(\gamma_m z)}{\gamma_m}\int_0^{\infty}\int_0^t f_{2m}(\xi,t-\tau)f_{3m}(\xi,\tau)d\tau d\xi, \tag{30}$$

where $\gamma_m = \frac{(2k-1)\pi}{2h}$. It is worth pointing out that the first term of Eq. (30), namely

$$v_N(y,z,t) = \frac{4f}{\rho\pi h}\sum_{k=1}^{\infty}\frac{(-1)^k \cos(\gamma_m z)}{\gamma_m}\int_0^{\infty} f_{1m}(\xi,t)\cos(y\xi)d\xi, \tag{31}$$

represents the complex velocity corresponding to a Newtonian fluid performing the same motion. Indeed, this term does not contain $\lambda$, $\lambda_r$, $\alpha$ and $\beta$ into its expression and the limit of $f_{3m}(\xi,t)$ from the second term of Eq. (30), is zero for $\lambda$ and $\lambda_r \to 0$. Consequently, the general solution $v(y,z,t)$, as it results from Eq. (30), is the sum between the Newtonian solution $v_N(y,z,t)$ and the non-Newtonian contribution $v_{nN}(y,z,t)$.

The inverse Laplace transform of $F_{1m}(\xi,q)$, namely

$$f_{1m}(\xi,t) = \frac{\nu(\xi^2+\gamma_m^2)-i\omega}{(\nu(\xi^2+\gamma_m^2))^2+\omega^2}\left[\cos(\omega t)+i\sin(\omega t)-e^{-\nu(\xi^2+\gamma_m^2)t}\right], \tag{32}$$

can be immidiately attained. Introducing Eq. (32) into (31) and using Eqs. (A2) - (A5) from Appendix, it results that

$$v_N(y,z,t) = \frac{2f}{\mu h}\sum_{k=1}^{\infty}\frac{(-1)^k\cos(\gamma_m z)}{\gamma_m}\frac{e^{-yB_m}}{\sqrt{A_m^2+B_m^2}}e^{i(\omega t-yA_m-\phi_m)} - \\ - \frac{4f}{\mu\pi h}\sum_{k=1}^{\infty}\frac{(-1)^k\cos(\gamma_m z)}{\gamma_m}\int_0^{\infty}\frac{(\xi^2+\gamma_m^2)-i\left(\dfrac{\omega}{\nu}\right)}{(\xi^2+\gamma_m^2)^2+\left(\dfrac{\omega}{\nu}\right)^2}e^{-\nu(\xi^2+\gamma_m^2)t}\cos(y\xi)d\xi, \tag{33}$$

where $2A_m^2 = \sqrt{\gamma_m^4+\dfrac{\omega^2}{\nu^2}} - \gamma_m^2$, $2B_m^2 = \sqrt{\gamma_m^4+\dfrac{\omega^2}{\nu^2}} + \gamma_m^2$ and $\tan\phi_m = \dfrac{A_m}{B_m}$. As expected, the starting solution (33) is a sum between the permanent and transient solutions.

In order to determine the non-Newtonian contribution $v_{nN}(y,z,t)$ of the complex velocity $v(y,z,t)$, we need the inverse Laplace transforms of the functions $F_{2m}(\xi,q)$ and $F_{3m}(\xi,q)$. For this, we firstly write them in the equivalent forms

$$F_{2m}(\xi,q) = q^{\beta-2}\left(1 - \left(\frac{\nu(\xi^2+\gamma_m^2)(\nu(\xi^2+\gamma_m^2)-i\omega)}{(\nu(\xi^2+\gamma_m^2))^2+\omega^2}\right)\frac{1}{q+\nu(\xi^2+\gamma_m^2)} - \right.$$
$$\left. -\left(\frac{\omega^2(\nu(\xi^2+\gamma_m^2)-i\omega)}{(\nu(\xi^2+\gamma_m^2))^2+\omega^2}\right)\frac{q}{q^2+\omega^2} - \left(\frac{\omega^2(\omega+i\nu(\xi^2+\gamma_m^2))}{(\nu(\xi^2+\gamma_m^2))^2+\omega^2}\right)\frac{\omega}{q^2+\omega^2}\right), \quad (34)$$

respectively (see also Eq. (A1) from Appendix)

$$F_{3m}(\xi,q) = \sum_{p=0}^{\infty}\sum_{j=0}^{p}\frac{(-1)^p p!}{(p-j)!j!}\left(\frac{\nu(\xi^2+\gamma_m^2)}{\lambda^\alpha}\right)^p \lambda_r^{\beta j}\left(\frac{q^{\alpha-\beta+\beta j-p}}{(q^\alpha+\lambda^{-\alpha})^{p+1}} + a\nu(\xi^2+\gamma_m^2)\frac{q^{\beta j-p-1}}{(q^\alpha+\lambda^{-\alpha})^{p+1}}\right). \quad (35)$$

Applying the inverse Laplace transform to Eqs. (34) and (35), it results that

$$f_{2m}(\xi,t) = \frac{t^{1-\beta}}{\Gamma(2-\beta)} - \left(\frac{\nu(\xi^2+\gamma_m^2)(\nu(\xi^2+\gamma_m^2)-i\omega)}{(\nu(\xi^2+\gamma_m^2))^2+\omega^2}\right)\frac{1}{\Gamma(2-\beta)}\int_0^t s^{1-\beta} e^{-\nu(\xi^2+\gamma_m^2)(t-s)}ds -$$
$$-\left(\frac{\omega^2(\nu(\xi^2+\gamma_m^2)-i\omega)}{(\nu(\xi^2+\gamma_m^2))^2+\omega^2}\right)\frac{1}{\Gamma(2-\beta)}\int_0^t s^{1-\beta}\cos(\omega(t-s))ds - \quad (36)$$
$$-\left(\frac{\omega^2(\omega+i\nu(\xi^2+\gamma_m^2))}{(\nu(\xi^2+\gamma_m^2))^2+\omega^2}\right)\frac{1}{\Gamma(2-\beta)}\int_0^t s^{1-\beta}\sin(\omega(t-s))ds$$

and

$$f_{3m}(\xi,t) = \sum_{p=0}^{\infty}\sum_{j=0}^{p}\frac{(-1)^p p!}{(p-j)!j!}\left(\frac{\nu(\xi^2+\gamma_m^2)}{\lambda^\alpha}\right)^p \times$$
$$\times \lambda_r^{\beta j}\left[G_{\alpha,\alpha-\beta+j\beta-p,p+1}\left(-\frac{1}{\lambda^\alpha},t\right) + a\nu(\xi^2+\gamma_m^2)G_{\alpha,j\beta-p-1,p+1}\left(-\frac{1}{\lambda^\alpha},t\right)\right]. \quad (37)$$

Consequently, the non-Newtonian contribution to the general solution is

$$v_{nN}(y,z,t) = \frac{4f}{\rho\pi h}\sum_{k=1}^{\infty}\frac{(-1)^{k+1}\cos(\gamma_m z)}{\gamma_m}\int_0^\infty\int_0^t f_{2m}(\xi,t-s)f_{3m}(\xi,s)\cos(y\xi)d\tau d\xi, \quad (38)$$

where $f_{2m}(\xi,t)$ and $f_{3m}(\xi,t)$ are defined in Eqs. (36) and (37).

Seperating real and imaginary parts of Eq. (30) and bearing in mind Eqs. (33) and (38), we find that

$$u_c(y,z,t) = u_{cN}(y,z,t) + u_{cnN}(y,z,t), \quad u_s(y,z,t) = u_{sN}(y,z,t) + u_{snN}(y,z,t), \quad (39)$$

where

$$u_{cN}(y,z,t) = \frac{2f}{\mu h}\sum_{k=1}^{\infty}\frac{(-1)^k\cos(\gamma_m z)}{\gamma_m}\frac{e^{-yB_m}}{\sqrt{A_m^2+B_m^2}}\cos(\omega t - yA_m - \phi_m) -$$
$$-\frac{4f}{\pi\mu h}\sum_{k=1}^{\infty}\frac{(-1)^k\cos(\gamma_m z)}{\gamma_m}\int_0^\infty\frac{(\xi^2+\gamma_m^2)e^{-\nu(\xi^2+\gamma_m^2)t}}{(\xi^2+\gamma_m^2)^2+\left(\frac{\omega}{\nu}\right)^2}\cos(y\xi)d\xi \quad (40)$$

$$u_{cnN}(y,z,t) = \frac{4f}{\rho\pi h}\sum_{k=1}^{\infty}\frac{(-1)^{k+1}\cos(\gamma_m z)}{\gamma_m}\int_0^\infty\int_0^t\left(\sum_{p=0}^{\infty}\sum_{j=0}^{P}\frac{(-1)^p p!}{(p-j)!j!}\left(\frac{\nu(\xi^2+\gamma_m^2)}{\lambda^\alpha}\right)^p \lambda_r^{\beta j}\times\right.$$

$$\times\left[G_{\alpha,\alpha-\beta+j\beta-p,p+1}\left(-\frac{1}{\lambda^\alpha},t-\tau\right)+a\nu(\xi^2+\gamma_m^2)G_{\alpha,j\beta-p-1,p+1}\left(-\frac{1}{\lambda^\alpha},t-\tau\right)\right]\times$$

$$\times\left[\frac{\tau^{1-\beta}}{\Gamma(2-\beta)}-\frac{1}{\Gamma(2-\beta)}\int_0^\tau s^{1-\beta}e^{-\nu(\xi^2+\gamma_m^2)(\tau-s)}ds+\right.$$

$$+\frac{\omega^2}{\nu^2}\frac{1}{(\xi^2+\gamma_m^2)^2+\left(\frac{\omega}{\nu}\right)^2}\frac{1}{\Gamma(2-\beta)}\int_0^\tau s^{1-\beta}e^{-\nu(\xi^2+\gamma_m^2)(\tau-s)}ds -$$

$$-\frac{\omega^2}{\nu}\frac{(\xi^2+\gamma_m^2)}{(\xi^2+\gamma_m^2)^2+\left(\frac{\omega}{\nu}\right)^2}\frac{1}{\Gamma(2-\beta)}\int_0^\tau s^{1-\beta}\cos(\omega(\tau-s))ds -$$

$$\left.-\frac{\omega^2}{\nu^2}\frac{\omega}{(\xi^2+\gamma_m^2)^2+\left(\frac{\omega}{\nu}\right)^2}\frac{1}{\Gamma(2-\beta)}\int_0^\tau s^{1-\beta}\sin(\omega(\tau-s))ds\right]\right)d\tau\cos(y\xi)d\xi$$

(41)

and

$$u_{sN}(y,z,t) = \frac{2f}{\mu h}\sum_{k=1}^{\infty}\frac{(-1)^k\cos(\gamma_m z)}{\gamma_m}\frac{e^{-yB_m}}{\sqrt{A_m^2+B_m^2}}\sin(\omega t - yA_m - \phi_m) +$$

$$+\frac{4f}{\pi\mu h}\frac{\omega}{\nu}\sum_{k=1}^{\infty}\frac{(-1)^k\cos(\gamma_m z)}{\gamma_m}\int_0^\infty\frac{e^{-\nu(\xi^2+\gamma_m^2)t}}{(\xi^2+\gamma_m^2)^2+\left(\frac{\omega}{\nu}\right)^2}\cos(y\xi)d\xi,$$

(42)

$$u_{snN}(y,z,t) = \frac{4f}{\rho\pi h}\sum_{k=1}^{\infty}\frac{(-1)^{k+1}\cos(\gamma_m z)}{\gamma_m}\int_0^{\infty}\int_0^t\left(\sum_{p=0}^{\infty}\sum_{j=0}^p\frac{(-1)^p p!}{(p-j)!j!}\left(\frac{\nu(\xi^2+\gamma_m^2)}{\lambda^\alpha}\right)^p \lambda_r^{\beta j}\right.\times$$

$$\times\left[G_{\alpha,\alpha-\beta+j\beta-p,p+1}(-\frac{1}{\lambda^\alpha},t-\tau)+a\nu(\xi^2+\gamma_m^2)G_{\alpha,j\beta-p-1,p+1}(-\frac{1}{\lambda^\alpha},t-\tau)\right]\times$$

$$\times\left[\frac{\omega}{\nu}\frac{(\xi^2+\gamma_m^2)}{(\xi^2+\gamma_m^2)^2+\left(\frac{\omega}{\nu}\right)^2}\right)\frac{1}{\Gamma(2-\beta)}\int_0^\tau s^{1-\beta}e^{-\nu(\xi^2+\gamma_m^2)(\tau-s)}ds+$$

$$+\frac{\omega^2}{\nu^2}\frac{\omega}{(\xi^2+\gamma_m^2)^2+\left(\frac{\omega}{\nu}\right)^2}\frac{1}{\Gamma(2-\beta)}\int_0^\tau s^{1-\beta}\cos(\omega(\tau-s))ds-$$

$$-\frac{\omega^2}{\nu}\frac{(\xi^2+\gamma_m^2)}{(\xi^2+\gamma_m^2)^2+\left(\frac{\omega}{\nu}\right)^2}\frac{1}{\Gamma(2-\beta)}\int_0^\tau s^{1-\beta}\sin(\omega(\tau-s))ds\left.\right]\right)d\tau\cos(y\xi)d\xi. \tag{43}$$

At once $u_c(y,z,t)$ and $u_s(y,z,t)$ are determined, the corresponding shear stresses $\tau_{1c}(y,z,t)$, $\tau_{1s}(y,z,t)$, $\tau_{2c}(y,z,t)$ and $\tau_{2s}(y,z,t)$ are obtained in the same way from Eqs. (5) and (6) with the initial conditions $\tau_1(y,z,0)=0$, $\tau_2(y,z,0)=0$ but their general expressions are not presented here.

4. **Special cases and a check of results**

The similar solutions for Maxwell and second grade fluids with fractional derivatives as well as those for ordinary fluids performing the same motion are immediately obtained from general solutions (30) or (39) for suitable values of material and fractional parameters $\lambda$ and $\lambda_r$, respectively $\alpha$ and $\beta$. For $\lambda_r = 0$ and $\beta = 1$, for instance, the solutions corresponding to fractional Maxwell fluids are easily attained but their expressions are not given here. However, for a check of general results, some special cases will be considered:

4.1. **Case $\alpha = \beta = 1$ (Ordinary Oldroyd-B fluids)**

By making $\alpha$ and $\beta \to 1$ into Eqs. (39), the solutions

$$u_{cOB}(y,z,t) = \frac{2f}{\mu h}\sum_{k=1}^{\infty}\frac{(-1)^k \cos(\gamma_m z)}{\gamma_m}\frac{e^{-yB_m}}{\sqrt{A_m^2+B_m^2}}\cos(\omega t - yA_m - \phi_m) -$$

$$-\frac{4f}{\pi\mu h}\sum_{k=1}^{\infty}\frac{(-1)^k \cos(\gamma_m z)}{\gamma_m}\int_0^{\infty}\frac{(\xi^2+\gamma_m^2)e^{-\nu(\xi^2+\gamma_m^2)t}}{(\xi^2+\gamma_m^2)^2+\left(\frac{\omega}{\nu}\right)^2}\cos(y\xi)d\xi +$$

$$+\frac{4f}{\rho\pi h}\sum_{k=1}^{\infty}\frac{(-1)^{k+1}\cos(\gamma_m z)}{\gamma_m}\int_0^{\infty}\int_0^t\left(\sum_{p=0}^{\infty}\sum_{j=0}^{p}\frac{(-1)^p p!}{(p-j)!j!}\left(\frac{\nu(\xi^2+\gamma_m^2)}{\lambda}\right)^p \lambda_r^j\times\right.$$

$$\times\left[G_{1,j-p,p+1}\left(-\frac{1}{\lambda},t-s\right)++a\nu(\xi^2+\gamma_m^2)G_{1,j-p-1,p+1}\left(-\frac{1}{\lambda},t-s\right)\right]\times$$

$$\times\left[1-\left(\frac{1-e^{-\nu(\xi^2+\gamma_m^2)s}}{\nu(\xi^2+\gamma_m^2)}\right)+\frac{\omega^2}{\nu^2}\frac{1}{(\xi^2+\gamma_m^2)^2+\left(\frac{\omega}{\nu}\right)^2}\left(\frac{1-e^{-\nu(\xi^2+\gamma_m^2)s}}{\nu(\xi^2+\gamma_m^2)}\right)-\right.$$

$$\left.\left.-\frac{\omega}{\nu}\frac{(\xi^2+\gamma_m^2)}{(\xi^2+\gamma_m^2)^2+\left(\frac{\omega}{\nu}\right)^2}\sin(\omega\tau)-\frac{\omega^2}{\nu^2}\frac{(1-\cos(\omega s))}{(\xi^2+\gamma_m^2)^2+\left(\frac{\omega}{\nu}\right)^2}\right]ds\cos(y\xi)d\xi\right) \quad (44)$$

and

$$u_{sOB}(y,z,t) = \frac{2f}{\mu h} \sum_{k=1}^{\infty} \frac{(-1)^k \cos(\gamma_m z)}{\gamma_m} \frac{e^{-yB_m}}{\sqrt{A_m^2 + B_m^2}} \sin(\omega t - yA_m - \phi_m) +$$

$$+ \frac{4f}{\pi \mu h} \frac{\omega}{\nu} \sum_{k=1}^{\infty} \frac{(-1)^k \cos(\gamma_m z)}{\gamma_m} \int_0^{\infty} \frac{e^{-\nu(\xi^2 + \gamma_m^2)t}}{(\xi^2 + \gamma_m^2)^2 + \left(\frac{\omega}{\nu}\right)^2} \cos(y\xi) d\xi +$$

$$+ \frac{4f}{\rho \pi h} \sum_{k=1}^{\infty} \frac{(-1)^{k+1} \cos(\gamma_m z)}{\gamma_m} \int_0^{\infty} \int_0^{t} \left( \sum_{p=0}^{\infty} \sum_{j=0}^{p} \frac{(-1)^p p!}{(p-j)! j!} \left(\frac{\nu(\xi^2 + \gamma_m^2)}{\lambda}\right)^p \lambda_r^j \times \right.$$

$$\times \left[ G_{1,j-p,p+1}\left(-\frac{1}{\lambda}, t-s\right) + a\nu(\xi^2 + \gamma_m^2) G_{1,j-p-1,p+1}\left(-\frac{1}{\lambda}, t-s\right) \right] \times$$

$$\times \left[ \frac{\omega}{\nu} \frac{(\xi^2 + \gamma_m^2)}{(\xi^2 + \gamma_m^2)^2 + \left(\frac{\omega}{\nu}\right)^2} \right) \left( \frac{1 - e^{-\nu(\xi^2 + \gamma_m^2)s}}{\nu(\xi^2 + \gamma_m^2)} \right) +$$

$$\left. + \frac{\omega^2}{\nu^2} \frac{\sin(\omega s)}{(\xi^2 + \gamma_m^2)^2 + \left(\frac{\omega}{\nu}\right)^2} - \frac{\omega^2}{\nu} \frac{(\xi^2 + \gamma_m^2)(1 - \cos(\omega s))}{(\xi^2 + \gamma_m^2)^2 + \left(\frac{\omega}{\nu}\right)^2} \right] ds \cos(y\xi) d\xi \quad (45)$$

corresponding to Oldroyd-B fluids are obtained.

As a check of results that have been obtained, Figs. 1 and 2 clearly show that the diagrams of velocities $u_{cOB}(y,z,t)$ and $u_{sOB}(y,z,t)$ at the middle of the channel are almost identical to those obtained in [25, Eqs. (30a) and (30b)] by a different technique. However, unlike the results from [25], our solutions are presented as a sum between the Newtonian solutions and the non-Newtonian contributions. For $\lambda_r = 0$ they reduce to the solutions corresponding to ordinary Maxwell fluids.

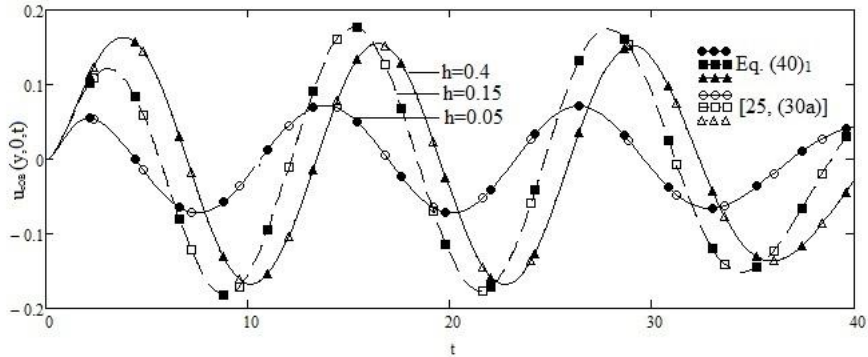

Fig. 1. Profile of velocities $u_{cOB}(y,0,t)$ given by Eq. (44) and [25, Eq. (30a)] with $f = -50$, $\nu = 0.001457$, $\rho = 1020$, $\lambda = 0.8$, $\lambda_r = 0.5$, $\omega = 0.5$, $y = 0.2$ and $h = 0.05, 0.15, 0.4$.

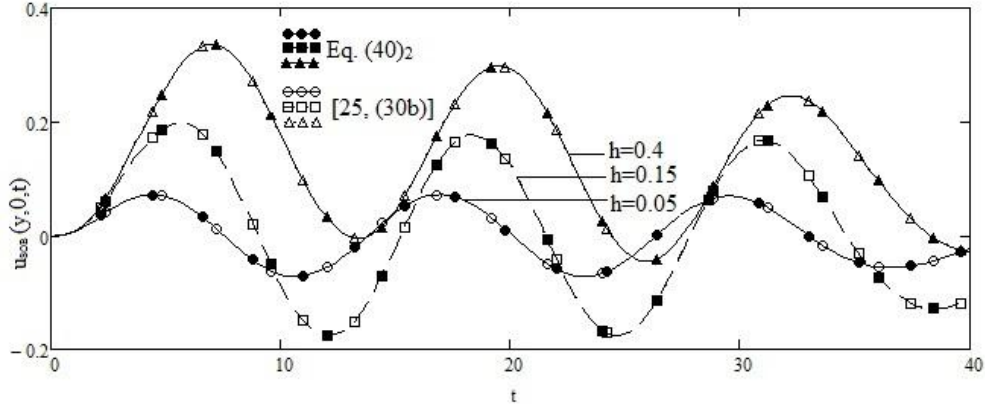

Fig. 2. Profile of velocities $u_{sOB}(y,0,t)$ given by Eq. (45) and [25, Eq. (30b)] with $f = -50$, $v = 0.001457$, $\rho = 1020$, $\lambda = 0.8$, $\lambda_r = 0.5$, $\omega = 0.5$, $y = 0.2$ and $h = 0.05, 0.15, 0.4$.

### 4.2.   Case $\lambda \to 0$ (fractional second grade fluids)

It is well known the fact that Oldroyd-B fluids do not contain second grade fluids as a special case. However, in some motions like those that have been here considered, the governing equations for Oldroyd-B fluids as well as those for fractional Oldroyd-B fluids remember the governing equations corresponding to second grade, respectively fractional second grade fluids. Consequently the solutions for fractional second grade fluids performing the same motion can be obtained as limiting cases of our solutions (39). Indeed, making the limit of Eqs. (39) for $\lambda \to 0$ and having in mind the known result

$$\lim_{\lambda \to 0} \frac{1}{\lambda^k} G_{a,b,k}\left(-\frac{c}{\lambda}, t\right) = \frac{1}{c^k} \frac{t^{-b-1}}{\Gamma(-b)}; b < 0, \qquad (46)$$

we find the solutions for fractional second grade fluids, namely

$$u_{cFSG}(y,z,t) = \frac{2f}{\mu h}\sum_{k=1}^{\infty}\frac{(-1)^k \cos(\gamma_m z)}{\gamma_m}\frac{e^{-yB_m}}{\sqrt{A_m^2+B_m^2}}\cos(\omega t - yA_m - \phi_m) -$$

$$-\frac{4f}{\pi\mu h}\sum_{k=1}^{\infty}\frac{(-1)^k \cos(\gamma_m z)}{\gamma_m}\int_0^{\infty}\frac{(\xi^2+\gamma_m^2)e^{-\nu(\xi^2+\gamma_m^2)t}}{(\xi^2+\gamma_m^2)^2+(\frac{\omega}{\nu})^2}\cos(y\xi)d\xi +$$

$$+\frac{4f}{\rho\pi h}\sum_{k=1}^{\infty}\frac{(-1)^{k+1}\cos(\gamma_m z)}{\gamma_m}\int_0^{\infty}\int_0^t\left(\sum_{p=0}^{\infty}\sum_{j=0}^{p}\frac{(-1)^p p!}{(p-j)!j!}\left(\nu(\xi^2+\gamma_m^2)\right)^{p+1}\lambda_r^{\beta(j+1)}\times\right.$$

$$\times\frac{(t-\tau)^{p-\beta j}}{\Gamma(p-j\beta+1)}\left[\frac{\tau^{1-\beta}}{\Gamma(2-\beta)}-\frac{1}{\Gamma(2-\beta)}\int_0^{\tau}s^{1-\beta}e^{-\nu(\xi^2+\gamma_m^2)(\tau-s)}ds+\right.$$

$$+\frac{\omega^2}{\nu^2}\frac{1}{(\xi^2+\gamma_m^2)^2+\left(\frac{\omega}{\nu}\right)^2}\frac{1}{\Gamma(2-\beta)}\int_0^{\tau}s^{1-\beta}e^{-\nu(\xi^2+\gamma_m^2)(\tau-s)}ds -$$

$$-\frac{\omega^2}{\nu}\frac{(\xi^2+\gamma_m^2)}{(\xi^2+\gamma_m^2)^2+\left(\frac{\omega}{\nu}\right)^2}\frac{1}{\Gamma(2-\beta)}\int_0^{\tau}s^{1-\beta}\cos(\omega(\tau-s))ds -$$

$$\left.\left.-\frac{\omega^2}{\nu^2}\frac{\omega}{(\xi^2+\gamma_m^2)^2+\left(\frac{\omega}{\nu}\right)^2}\frac{1}{\Gamma(2-\beta)}\int_0^{\tau}s^{1-\beta}\sin(\omega(\tau-s))ds\right]\right)d\tau\cos(y\xi)d\xi, \quad (47)$$

$$u_{sFSG}(y,z,t) = \frac{2f}{\mu h}\sum_{k=1}^{\infty}\frac{(-1)^k \cos(\gamma_m z)}{\gamma_m}\frac{e^{-yB_m}}{\sqrt{A_m^2+B_m^2}}\sin(\omega t - yA_m - \phi_m) +$$

$$+ \frac{4f}{\pi\mu h}\frac{\omega}{\nu}\sum_{k=1}^{\infty}\frac{(-1)^k \cos(\gamma_m z)}{\gamma_m}\int_0^{\infty}\frac{e^{-\nu(\xi^2+\gamma_m^2)t}}{(\xi^2+\gamma_m^2)^2 + \left(\frac{\omega}{\nu}\right)^2}\cos(y\xi)d\xi +$$

$$+ \frac{4f}{\rho\pi h}\sum_{k=1}^{\infty}\frac{(-1)^{k+1}\cos(\gamma_m z)}{\gamma_m}\int_0^{\infty}\int_0^t\left(\sum_{p=0}^{\infty}\sum_{j=0}^{p}\frac{(-1)^p p!}{(p-j)!j!}\left(\nu(\xi^2+\gamma_m^2)\right)^{p+1}\lambda_r^{\beta(j+1)}\times\right.$$

$$\times\frac{(t-\tau)^{p-j\beta}}{\Gamma(p-j\beta+1)}\left[\frac{\omega}{\nu}\frac{(\xi^2+\gamma_m^2)}{(\xi^2+\gamma_m^2)^2+\left(\frac{\omega}{\nu}\right)^2}\right)\frac{1}{\Gamma(2-\beta)}\int_0^{\tau}s^{1-\beta}e^{-\nu(\xi^2+\gamma_m^2)(\tau-s)}ds +$$

$$+\frac{\omega^2}{\nu^2}\frac{\omega}{(\xi^2+\gamma_m^2)^2+\left(\frac{\omega}{\nu}\right)^2}\frac{1}{\Gamma(2-\beta)}\int_0^{\tau}s^{1-\beta}\cos(\omega(\tau-s))ds -$$

$$\left.-\frac{\omega^2}{\nu}\frac{(\xi^2+\gamma_m^2)}{(\xi^2+\gamma_m^2)^2+\left(\frac{\omega}{\nu}\right)^2}\frac{1}{\Gamma(2-\beta)}\int_0^{\tau}s^{1-\beta}\sin(\omega(\tau-s))ds\right]\right)d\tau\cos(y\xi)d\xi. \quad (48)$$

Now, by making $\beta=1$ into Eqs. (47) and (48), the solutions corresponding to second grade fluids, namely

$$u_{cSG}(y,z,t) = \frac{2f}{\mu h}\sum_{k=1}^{\infty}\frac{(-1)^k \cos(\gamma_m z)}{\gamma_m}\frac{e^{-yB_m}}{\sqrt{A_m^2+B_m^2}}\cos(\omega t - yA_m - \phi_m) -$$

$$- \frac{4f}{\pi\mu h}\sum_{k=1}^{\infty}\frac{(-1)^k \cos(\gamma_m z)}{\gamma_m}\int_0^{\infty}\frac{(\xi^2+\gamma_m^2)e^{-\nu(\xi^2+\gamma_m^2)t}}{(\xi^2+\gamma_m^2)^2+\left(\frac{\omega}{\nu}\right)^2}\cos(y\xi)d\xi +$$

$$+ \frac{4f}{\rho\pi h}\sum_{k=1}^{\infty}\frac{(-1)^{k+1}\cos(\gamma_m z)}{\gamma_m}\int_0^{\infty}\int_0^t\left(\sum_{p=0}^{\infty}\sum_{j=0}^{p}\frac{(-1)^p p!}{(p-j)!j!}\left(\nu(\xi^2+\gamma_m^2)\right)^{p+1}\lambda_r^{j+1}\times\right.$$

$$\times\frac{(t-\tau)^{p-j}}{\Gamma(p-j+1)}\left[1+\frac{1}{\nu}\frac{e^{-\nu(\xi^2+\gamma_m^2)(\tau-s)}-1}{(\xi^2+\gamma_m^2)^2+\left(\frac{\omega}{\nu}\right)^2}-\frac{\omega}{\nu}\frac{(\xi^2+\gamma_m^2)\sin(\omega\tau)}{(\xi^2+\gamma_m^2)^2+\left(\frac{\omega}{\nu}\right)^2}-\right.$$

$$\left.\left.-\frac{\omega^2}{\nu^2}\frac{1-\cos(\omega\tau)}{(\xi^2+\gamma_m^2)^2+\left(\frac{\omega}{\nu}\right)^2}\right]\right)d\tau\cos(y\xi)d\xi, \quad (49)$$

and

$$u_{sSG}(y,z,t) = \frac{2f}{\mu h}\sum_{k=1}^{\infty}\frac{(-1)^k \cos(\gamma_m z)}{\gamma_m}\frac{e^{-yB_m}}{\sqrt{A_m^2+B_m^2}}\sin(\omega t - yA_m - \phi_m) +$$

$$+\frac{4f}{\pi\mu h}\frac{\omega}{v}\sum_{k=1}^{\infty}\frac{(-1)^k \cos(\gamma_m z)}{\gamma_m}\int_0^{\infty}\frac{e^{-v(\xi^2+\gamma_m^2)t}}{(\xi^2+\gamma_m^2)^2+\left(\frac{\omega}{v}\right)^2}\cos(y\xi)d\xi +$$

$$+\frac{4f}{\rho\pi h}\sum_{k=1}^{\infty}\frac{(-1)^{k+1}\cos(\gamma_m z)}{\gamma_m}\int_0^{\infty}\int_0^{t}\left(\sum_{p=0}^{\infty}\sum_{j=0}^{p}\frac{(-1)^p p!}{(p-j)!j!}\left(v(\xi^2+\gamma_m^2)\right)^{p+1}\lambda_r^{j+1}\times\right.$$

$$\times\frac{(t-\tau)^{p-j}}{\Gamma(p-j+1)}\left[\frac{\omega}{v^2}\frac{1-e^{-v(\xi^2+\gamma_m^2)\tau}}{(\xi^2+\gamma_m^2)^2+\left(\frac{\omega}{v}\right)^2}+\frac{\omega^2}{v^2}\frac{\sin(\omega\tau)}{(\xi^2+\gamma_m^2)^2+\left(\frac{\omega}{v}\right)^2}-\right.$$

$$\left.\left.-\frac{\omega}{v}\frac{(\xi^2+\gamma_m^2)(1-\cos(\omega\tau))}{(\xi^2+\gamma_m^2)^2+\left(\frac{\omega}{v}\right)^2}\right]d\tau\cos(y\xi)d\xi. \qquad (50)$$

are recovered.

Indeed, Figs. 3 amd 4 clearly show that the diagrams of velocities $u_{cSG}(y,z,t)$ and $u_{sSG}(y,z,t)$ at the middle of the channel are almost identical to those obtained in [29, Eqs. (22) and (23)] by a different technique.

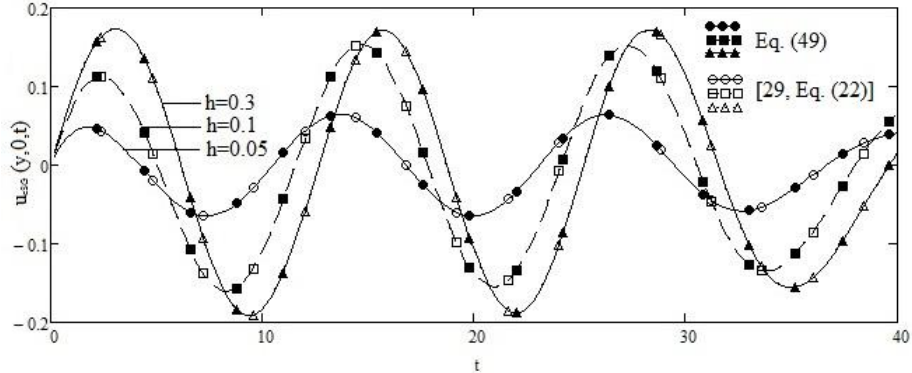

Fig. 3. Profile of velocities $u_{cSG}(y,0,t)$ given by Eq. (49) and [29, Eq. (22)] with $f=-50$, $v=0.001457$, $\rho=1020$, $\lambda=0.8$, $\lambda_r=0.5$, $\omega=0.5$, $y=0.2$ and $h=0.05,0.1,0.3$.

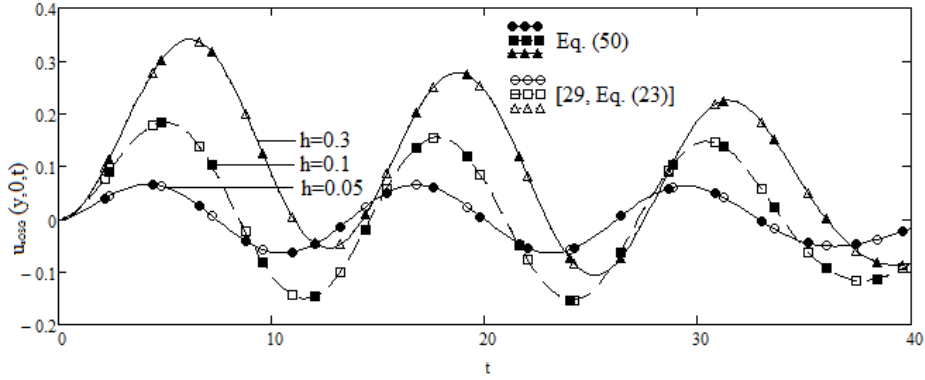

Fig. 4. Profile of velocities $u_{sSG}(y,0,t)$ given by Eq. (50) and [29, Eq. (23)] with $f=-50$, $v=0.001457$, $\rho=1020$, $\lambda=0.8$, $\lambda_r=0.5$, $\omega=0.5$, $y=0.2$ and $h=0.05, 0.1, 0.3$.

For $\lambda_r=0$, as expected, Eqs. (49) and (50) reduce to the Newtonian solutions that are identical to those obtained in [30, Eqs. (14) and (18)].

## 5. Flow over an infinite plate

In the absence of side walls, namely when $h\to\infty$, Eqs. (39) take the simpler forms (see also Eq. (A6) from Appendix)

$$v_c(y,t) = -\frac{f}{\mu}\sqrt{\frac{v}{\omega}}e^{-y\sqrt{\frac{\omega}{2v}}}\cos(\omega t - y\sqrt{\frac{\omega}{2v}} - \frac{\pi}{4}) + \frac{2f}{\pi\mu}\int_0^\infty \frac{\xi^2 e^{-v\xi^2 t}}{\xi^4 + \left(\frac{\omega}{v}\right)^2}\cos(y\xi)d\xi +$$

$$+\frac{2f}{\rho\pi}\int_0^\infty\int_0^t\left(\sum_{p=0}^\infty\sum_{j=0}^p \frac{(-1)^p p!}{(p-j)!j!}(\frac{v\xi^2}{\lambda^\alpha})^p \lambda_r^{\beta j}\left[G_{\alpha,\alpha-\beta+j\beta-p,p+1}(-\frac{1}{\lambda^\alpha},t-\tau)+\right.\right.$$

$$\left.+av\xi^2 G_{\alpha,j\beta-p-1,p+1}(-\frac{1}{\lambda^\alpha},t-\tau)\right]\times\left[\frac{\tau^{1-\beta}}{\Gamma(2-\beta)}-\frac{1}{\Gamma(2-\beta)}\int_0^\tau s^{1-\beta}e^{-v\xi^2(\tau-s)}ds +$$

$$+\left(\frac{\omega}{v}\right)^2\frac{1}{\xi^4+\left(\frac{\omega}{v}\right)^2}\frac{1}{\Gamma(2-\beta)}\int_0^\tau s^{1-\beta}e^{-v\xi^2(\tau-s)}ds -$$

$$-\frac{\omega^2}{v}\frac{\xi^2}{\xi^4+\left(\frac{\omega}{v}\right)^2}\frac{1}{\Gamma(2-\beta)}\int_0^\tau s^{1-\beta}\cos(\omega(\tau-s))ds -$$

$$\left.-\left(\frac{\omega}{v}\right)^2\frac{\omega}{\xi^4+\left(\frac{\omega}{v}\right)^2}\frac{1}{\Gamma(2-\beta)}\int_0^\tau s^{1-\beta}\sin(\omega(\tau-s))ds\right]\right)d\tau\cos(y\xi)d\xi \quad (51)$$

and

$$v_s(y,t) = -\frac{f}{\mu}\sqrt{\frac{\nu}{\omega}}e^{-y\sqrt{\frac{\omega}{2\nu}}}\sin(\omega t - y\sqrt{\frac{\omega}{2\nu}} - \frac{\pi}{4}) - \frac{2f}{\pi\mu}\frac{\omega}{\nu}\int_0^\infty \frac{e^{-\nu\xi^2 t}}{\xi^4 + (\frac{\omega}{\nu})^2}\cos(y\xi)d\xi +$$

$$+\frac{2f}{\rho\pi}\int_0^\infty\int_0^t \left(\sum_{p=0}^\infty\sum_{j=0}^p \frac{(-1)^p p!}{(p-j)!j!}(\frac{\nu\xi^2}{\lambda^\alpha})^p \lambda_r^{\beta j} \times \right.$$

$$\times\left[ G_{\alpha,\alpha-\beta+j\beta-p,p+1}(-\frac{1}{\lambda^\alpha}, t-\tau) + a\nu\xi^2 G_{\alpha,j\beta-p-1,p+1}(-\frac{1}{\lambda^\alpha}, t-\tau)\right]\times$$

$$\times\left[\frac{\omega}{\nu}\frac{\xi^2}{\xi^4+(\frac{\omega}{\nu})^2}\right)\frac{1}{\Gamma(2-\beta)}\int_0^\tau s^{1-\beta}e^{-\nu\xi^2(\tau-s)}ds + $$

$$+\frac{\omega^2}{\nu^2}\frac{\omega}{\xi^4+(\frac{\omega}{\nu})^2}\frac{1}{\Gamma(2-\beta)}\int_0^\tau s^{1-\beta}\cos(\omega(\tau-s))ds - $$

$$\left.\left.-\frac{\omega^2}{\nu}\frac{\xi^2}{\xi^4+(\frac{\omega}{\nu})^2}\frac{1}{\Gamma(2-\beta)}\int_0^\tau s^{1-\beta}\sin(\omega(\tau-s))ds\right]\right)d\tau\cos(y\xi)d\xi \qquad (52)$$

corresponding to the motion of an incompressible fractional Oldroyd-B fluid over an infinite plate. By making $\alpha, \beta \to 1$ into Eqs. (51) and (52), the solutions

$$v_c(y,t) = -\frac{f}{\mu}\sqrt{\frac{\nu}{\omega}}e^{-y\sqrt{\frac{\omega}{2\nu}}}\cos(\omega t - y\sqrt{\frac{\omega}{2\nu}} - \frac{\pi}{4}) + \frac{2f}{\pi\mu}\int_0^\infty \frac{\xi^2 e^{-\nu\xi^2 t}}{\xi^4 + \left(\frac{\omega}{\nu}\right)^2}\cos(y\xi)d\xi +$$

$$+\frac{2f}{\rho\pi}\int_0^\infty\int_0^t \left(\sum_{p=0}^\infty\sum_{j=0}^p \frac{(-1)^p p!}{(p-j)!j!}(\frac{\nu\xi^2}{\lambda})^p \lambda_r^j \left[G_{1,j-p,p+1}(-\frac{1}{\lambda}, t-\tau) + \right.\right.$$

$$+a\nu\xi^2 G_{1,j-p-1,p+1}(-\frac{1}{\lambda}, t-\tau)\right]\times$$

$$\left.\times\left[1+\frac{1}{\nu}\frac{\xi^2(1-e^{-\nu\xi^2\tau})}{\xi^4+\left(\frac{\omega}{\nu}\right)^2}-\frac{\omega}{\nu}\frac{\xi^2\sin(\omega\tau)}{\xi^4+\left(\frac{\omega}{\nu}\right)^2}-\left(\frac{\omega}{\nu}\right)^2\frac{1-\cos(\omega\tau)}{\xi^4+\left(\frac{\omega}{\nu}\right)^2}\right]\right)d\tau\cos(y\xi)d\xi \qquad (53)$$

and

$$v_s(y,t) = -\frac{f}{\mu}\sqrt{\frac{v}{\omega}}e^{-y\sqrt{\frac{\omega}{2v}}}\sin(\omega t - y\sqrt{\frac{\omega}{2v}} - \frac{\pi}{4}) - \frac{2f}{\pi\mu}\frac{\omega}{v}\int_0^\infty \frac{e^{-v\xi^2 t}}{\xi^4 + \left(\frac{\omega}{v}\right)^2}\cos(y\xi)d\xi +$$

$$+\frac{2f}{\rho\pi}\int_0^\infty\int_0^t\left(\sum_{p=0}^\infty\sum_{j=0}^p \frac{(-1)^p p!}{(p-j)!j!}(\frac{v\xi^2}{\lambda})^p \lambda_r^j \times\right.$$

$$\times\left[G_{1,j-p,p+1}(-\frac{1}{\lambda},t-\tau) + av\xi^2 G_{1,j-p-1,p+1}(-\frac{1}{\lambda},t-\tau)\right]\times \qquad (54)$$

$$\times\left[\frac{\omega}{v^2}\frac{1-e^{-v\xi^2\tau}}{\xi^4+\left(\frac{\omega}{v}\right)^2} + \frac{\omega^2}{v^2}\frac{\sin(\omega\tau)}{\xi^4+\left(\frac{\omega}{v}\right)^2} - \frac{\omega}{v}\frac{\xi^2(1-\cos(\omega\tau))}{\xi^4+\left(\frac{\omega}{v}\right)^2}\right]d\tau\cos(y\xi)d\xi,$$

corresponding to an Oldroyd-B fluid performing the same motion are recovered. Indeed, Fig. 5 clearly show that profiles of the velocities $v_c(y,t)$ and $v_s(y,t)$ given by Eqs. (53) and (54) are almost identical to those obtained in [33, Eqs. (27) and (28)].

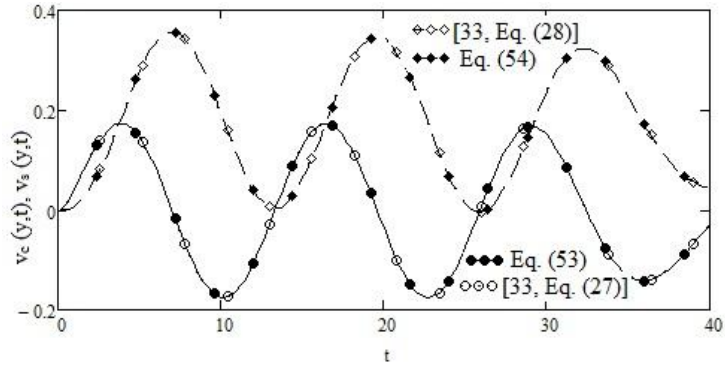

**Fig. 5.** Profile of velocities $v_c(y,t)$ given by Eq. (53) and [33, Eq. (27)], $v_s(y,t)$ given by Eq. (54) and [33, Eq. (28)] versus $t$ with $f=-50$, $v=0.001457$, $\rho=1020$, $\lambda=0.8$, $\lambda_r=0.5$, $\omega=0.5$, $y=0.02$.

## 6. Numerical results and discussions

In this note, two types of unsteady motions of fractional Oldroyd-B fluids between two parallel walls perpendicular to a plate are studied by means of integral transforms. Exact solutions are developed for velocity in terms of the generalized $G_{a,b,c}(d,.)$ functions and they are presented as a sum between Newtonian solutions and the non-Newtonian contributions. This is important because the non-Newtonian behavior of the fluid motion is brought to light. Furthermore, the solutions corresponding to second grade and Maxwell fluids with fractional derivatives as well as those for ordinary fluids performing the same motion are obtained as limiting cases of general solutions. As a check of their validity, the profiles of velocity field given by Eqs. (44), (45), (49) and (50) as well as those of the similar solutions from the existing literature have been presented in Figs. 1-4 for $y=0$ and three different values of the distance $h$ between walls. They clearly show that our limiting solutions for Oldroyd-B and second grade fluids are equivalent to the known solutions from the literature.

In the absence of the side walls, namely when the distance between walls tends to infinity, all solutions are going to the similar solutions corresponding to the motion over an infinite plate. These solutions can be also particularized to give the similar solutions for ordinary or fractional Maxwell and second grade fluids. The solutions corresponding to ordinary Oldroyd-B fluids, for instance, are given by Eqs. (53) and (54). Their diagrams, as it results from Fig. 5, are almost identical to those corresponding to the solutions (27) and (28) that have been obtained by Shahid et al. [33] by a different technique.

In order to determine the distance between walls for which the measured value of velocity in the middle of the channel is unaffected by the presence of the side walls (namely, this value it is approximately equal to the velocity corresponding to the motion over an infinite plate), Figs. 6 and 7 have been sketched for three distinct values of the time $t$. It is observed that this distance for cosine oscillations of the shear ($h_c = 0.5$) is smaller than that for the sine oscillations ($h_s = 0.76$). This is obvious, because, at time $t = 0^+$ the shear stress on the boundary is zero in the second case.

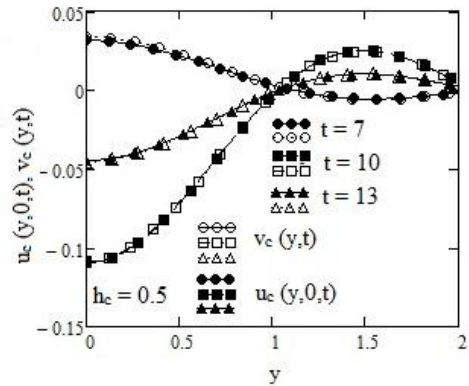

Fig. 6. Profile of velocities $u_c(y,0,t)$ given by Eq. (39)$_1$ and $v_c(y,t)$ given by Eq. (51) for $f = -50$, $v = 0.001457$, $\lambda = 2$, $\lambda_r = 1.5$, $\alpha = 0.8$, $\beta = 0.4$, $\omega = 0.5$ and different values of $t$.

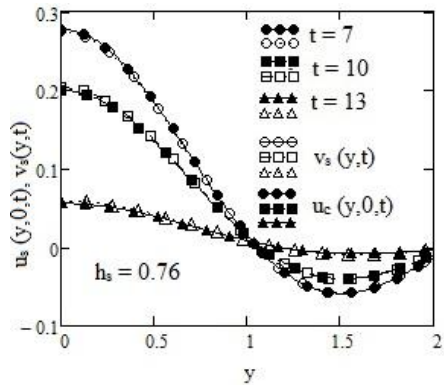

Fig. 7. Profile of velocities $u_s(y,0,t)$ given by Eq. (39)$_2$ and $v_s(y,t)$ given by Eq. (52) for $f = -50$,

$v = 0.001457$, $\lambda = 2$, $\lambda_r = 1.5$, $\alpha = 0.8$, $\beta = 0.4$, $\omega = 0.5$ and different values of $t$.

In order to avoid repetition only the influence of fractional parameters $\alpha$ and $\beta$ on the fluid motion (effects of the material parameters have been shown in [25]) is underlined by Figs. 8 and 9. As expected, they have opposite effects on the fluid motion in both cases. For a fixed value of $\beta$ an increase of the parmeter $\alpha$ implies a faster flow of the fluid in the case of cosine oscillations and a slowness of the flow in the second case. Opposite effects appear when the parameter $\alpha$ is unchanging and $\beta$ increases. In all cases the fluid velocity decreases from a maximum value near the bottom plate and continuously decreases for large values of $y$.

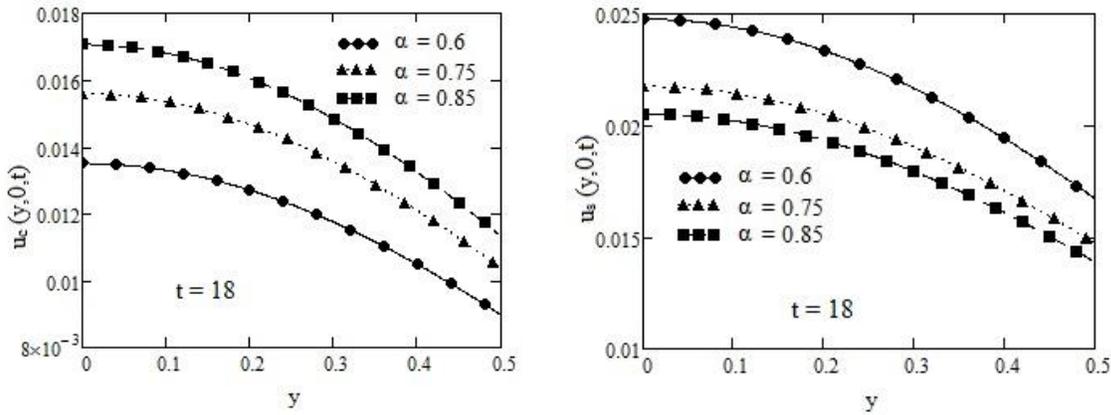

Fig. 8. Profile of velocities $u_c(y,0,t)$ given by Eq. (39)$_1$ and $u_s(y,0,t)$ given by Eq. (39)$_2$ for $f = -50$, $v = 0.001457$, $\lambda = 0.8$, $\lambda_r = 0.5$, $\beta = 0.4$, $\omega = 1.5$ at $t = 18$ and different values of $\alpha$.

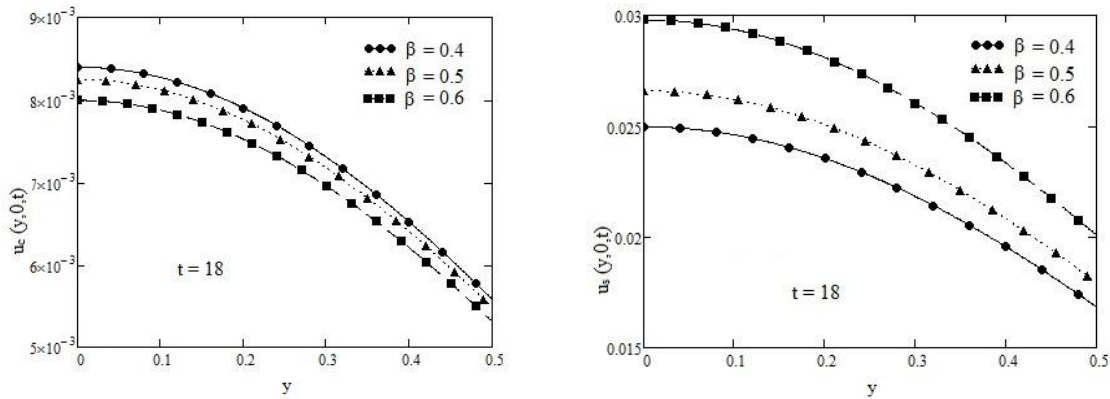

Fig. 9. Profile of velocities $u_c(y,0,t)$ given by Eq. (39)$_1$ and $u_s(y,0,t)$ given by Eq. (39)$_2$ for $f = -50$, $v = 0.001457$, $\lambda = 0.8$, $\lambda_r = 0.5$, $\alpha = 0.85$, $\omega = 1.5$ at $t = 18$ and different values of $\beta$.

Appendix

$$\frac{1}{x+a} = \sum_{k=0}^{\infty}(-1)^k \frac{x^k}{a^{k+1}}, (a+b)^k = \sum_{m=0}^{k}\frac{k!}{(k-m)!m!}a^m b^{k-m}, \tag{A1}$$

$$\int_0^{\infty}\frac{(\xi^2+d^2)\cos(y\xi)}{(\xi^2+d^2)^2+c^2}d\xi = \frac{\pi e^{-yB}}{2(A^2+B^2)}\left[B\cos(yA)-A\sin(yA)\right], \tag{A2}$$

$$\int_0^{\infty}\frac{\cos(y\xi)}{(\xi^2+d^2)^2+c^2}d\xi = \frac{\pi e^{-yB}}{2c(A^2+B^2)}\left[A\cos(yA)+B\sin(yA)\right], \tag{A3}$$

$$\int_0^{\infty}\frac{(\xi^2+d^2)\xi\sin(y\xi)}{(\xi^2+d^2)^2+c^2}d\xi = \frac{\pi}{2}e^{-yB}\cos(yA), \tag{A4}$$

$$\int_0^{\infty}\frac{\xi\sin(y\xi)}{(\xi^2+d^2)^2+c^2}d\xi = \frac{\pi}{2c}e^{-yB}\sin(yA), \tag{A5}$$

where $2A^2 = \sqrt{d^4+c^2}-d^2$, $2B^2 = \sqrt{d^4+c^2}+d^2$

$$\lim_{h\to\infty}\frac{2}{h}\sum_{n=1}^{\infty}\frac{(-1)^{n+1}\cos(\gamma_m z)}{\gamma_m}=1. \tag{A6}$$